\documentclass[12pt,preprint,showkeys]{revtex4-2}
\usepackage{graphicx}
\usepackage{amsmath}
\usepackage{amsfonts,amssymb}
\usepackage[matrix,arrow,curve,frame,poly,arc]{xy}
\usepackage[english]{babel}

\def\be{\begin{equation}}
\def\ee{\end{equation}}
\def\ba{\begin{eqnarray}}
\def\ea{\end{eqnarray}}

\begin{document}
\bibliographystyle{plainnat}

\title{Scheme of integration of vacuum $ F(R) $ gravity in a travelling wave variable.}

\author{Maria V. Shubina}

\email{yurova-m@rambler.ru}

\affiliation{Skobeltsyn Institute of Nuclear Physics\\Lomonosov Moscow State University
\\ Leninskie gory, GSP-1, Moscow 119991, Russian Federation}


\begin{abstract}

In this article we propose the scheme of integration of two-dimensional $F(R)$ gravity vacuum equations in a travelling wave variable. The main emphasis is placed on the fundamental possibility of obtaining different forms of the function $F(R)$ by arbitrarily choosing a certain function through which all components of the metric tensor of the theory can be expressed.

\end{abstract}


\maketitle
\newpage

\section{Introduction}

Despite the fact that to this day Einstein’s General Theory of Relativity (GTR) remains the most reliable and consistent fundamental theory describing gravitational interaction, in recent decades modifications of GTR have appeared that make it possible to correctly describe the cosmological dynamic, the transition between different cosmological epochs of the evolution of the Universe and dynamics of cosmological perturbations. General Relativity is described by the Einstein-Hilbert gravitational action with Lagrange density $ \sqrt{-g} R $ where $ R $ is the Ricci scalar, or the Ricci curvature ($ g = \det g_{\mu \nu }$). One of the natural generalizations of the General Theory of Relativity is the most popular $F(R)$ gravity in which the Lagrangian is considered as a function of the Ricci scalar, see \cite{SF_2010}-\cite{NOO} and references therein. Numerous authors consider various forms of the function $F(R)$ such as $ F(R) \sim R + f(R) $ where $ f(R) $ may have a rather arbitrary form, in particular $ f(R) \sim (R + c)^k$, $ f(R) \sim  a + b e^{cR} $, $ f(R) \sim R \ln R $ and others; models with $ F(R) \sim (a + R)^n  $, $ F(R) \sim R^n +a R^m $ and others are also considered and the corresponding properties of space-time and the objects it describes are analysed \cite{CB2006}-\cite{AZM2014}. In the general case the $F(R)$ gravity equations turn out to be very complicated and in most works the authors find exact solutions for the case of dependence of theory variables on one or, less often, two coordinates. This is also due to the fact that the metrics of interest from the point of view of cosmology, understanding the nature of the evolution of the Universe, are mostly diagonal and depend on $ t $ (or on $ t, r $); the same can be said about objects such as black holes, wormholes and others, whose metric depends also on one or two coordinates. Therefore the authors obtain exact analytical and numerical solutions describing singular and non-singular black holes, traversable wormholes coupled with and without a matter field mainly for the dependence of the metric on a single variable and with a predefined function $ F(R) $ (see the references above). 

In this paper it seemed interesting to us to move away from the analysis of the properties of objects and metrics obtained for the initially specified $F(R) $ and to focus specifically on the possibility of obtaining different forms of the function $ F(R) $. The derivation of exact solutions in $ F(R) $ gravity cosmological models is actively developing. In the work \cite{P2016} Paliathanasis consider $F(R) $ gravity in a Friedmann-Lema\^itre-Robertson-Walker (FLRW) spacetime and apply the Killing tensors of the minisuperspace in order to define the forms of $F(R) $ and the field equations that admit quadratic conservation laws given by Noether’s Theorem. The author finds three new integrable $F(R) $ models for which he reduces the field equations to a system of two nonlinear first order ordinary differential equations. The singularity analysis has been applied to the case of $ F(R) = R + q R^n $ gravity \cite{PL2016}. A number of exact time-dependent solutions including $ F(R) \sim R^2 +\beta R + \gamma$ were obtained by Vernov \textit{et al.} \cite{VIP2020} using the superpotential method. In \cite{IV2021} the modified gravity cosmological models that can be transformed into two-field chiral cosmological models by the conformal metric transformation are considered. For these models the authors have found general solutions in the cosmic time that corresponds to a parametric time in the Jordan frame and this makes it possible to get the general solutions for the corresponding modified gravity models in parametric time. Using this method Ivanov \textit{et al.} have found general solutions for the $ F(R) \sim R^2 $ gravity in the FLRW metric. In \cite{MSh_1_2024} using a certain ansatz on the Hubble rate a number of exact analytical solutions of vacuum $ F(R) $ gravity depending on one variables are obtained. 

Also many authors study solutions that depend on one radial coordinate and describe different black holes \cite{N2021}-\cite{KPTW2021} and references therein. In \cite{N2021} statically rotating uncharged anti-de-Sitter (AdS) and de-Sitter (dS) black holes were obtained and studied in detail within the framework of the $ F(R) $ theory without assuming any constraints on $ F(R) $. Considering a spherically symmetric metric ansatz and without specifying the form of $ F(R) $ a general charged black hole solution in $ D $ dimensions was found in \cite{PTW2021}. In the work \cite{KPTW2021} Karakasis \textit{et al.} derived asymptotically flat or (A)dS exact black hole solutions with dynamic Ricci curvature. The authors consider the asymptotically flat model $ F(R) = R - 2 \alpha \sqrt{R} $ and the asymptotically (A)dS model $ F(R) = R - 2 \Lambda - 2 \alpha \sqrt{R - 4 \Lambda} $ where $ \alpha \neq 0 $.

As far as we know the methods described above have not been applied to the case of a theory dependent on two variables and this may be a topic for future research. Some exact solutions of $F(R)$ gravity depending on two variables were found in the works \cite{GPLM2024}-\cite{LLYXZ2024} and references therein. Now we propose a scheme for integrating two-dimensional $F(R)$ gravity equations in a travelling wave variable $ y $ for an arbitrary function $F(y)$, however our goal is still to obtain an exact analytical dependence $ F = F(R) $. Introducing for convenience of computation a certain function $ \chi(y) $ we obtain that the resulting equations allow to find exact expressions for all variables of the theory as functions of $\chi(y)$ and this function $\chi(y)$ we set at our discretion. But it is difficult to see in advance whether $ y $ can be expressed as a function of $ R $; and even if it is possible to express $y=y(R)$ it is not obvious that this function has an inverse one, that is, that it is possible to find $ R = R(y) $. Therefore we impose on $ \chi(y) $ the condition of its dependence on $R$, which allows to obtain exactly $F(R)$ as a function of scalar curvature. We give several application examples of this scheme in which $ F(R) $ is obtained both exactly and in quadratures. Metric functions are also expressed in terms of curvature, so a standard analysis of their physical properties is difficult and requires further investigation. However, since, as far as we know, our scheme is new, we believe that such a consideration of $F(R)$-gravity may be of interest to researchers, it may initiate the consideration of new forms of the $F(R)$ and may also be useful.

\section{Models under consideration and field equations}

In this paper we consider the metric $F(R)$-gravity model for the case when the metric tensor depends on two variable. The gravitational action without matter fields is 
\be
S = \frac{1}{16 \pi \kappa} \int d^{4}x \sqrt{-\textit{g}} \,F(R),
\ee
where $ \kappa $ is the gravitational constant, $ \textit{g} $ is the determinant of the metric tensor $ g_{\mu\nu} $, $ R = g^{\mu\nu} R_{\mu\nu} $ is the scalar curvature, or the Ricci scalar, $ R_{\mu\nu} $ is the Ricci tensor. Variation of eq.(1) with respect to the metric gives the field equations \cite{SF_2010}
\be
F_{R}(R) R_{\mu \nu} - \frac{1}{2}F(R)g_{\mu \nu} - [ \nabla_{\mu} \nabla_{\nu} - g_{\mu \nu} \square ] F_{R}(R) =  0,
\ee
where $ F_{R}(R)\equiv \dfrac{dF(R)}{dR} $.

We will consider a metric depending on two coordinates and take the metric interval in the form:
\be
ds^{2} = - 4 f(\zeta, \eta) \, d\zeta d\eta - g_{ab} dx^{a}dx^{b},
\ee
where $ g_{ab} = g_{ab} (\zeta, \eta) $, $ a, b = 1, 2 $ and the signature ($ + - - - $). This form of interval occurs in Einstein's theory of gravity, for example in the work \cite{BZ} and, in particular, the diagonal metric belongs to it. A detailed analysis of general metrics like eq. (3) is performed by Stephani \textit{et al} in \cite{SKMHH}; following \cite{SKMHH} we parametrize (up to notation) the components of the two-dimensional metric tensor $g_{ab}(\zeta, \eta)$ as follows:
\ba
g_{ab} = \begin{pmatrix}
\psi & \psi \, \omega\\
\psi \, \omega & \psi \, {\omega}^2 + \sigma_{\alpha}^{2} \,\, \alpha^{2}\, \psi^{-1}
\end{pmatrix}, \,\,\, \sigma_{\alpha}^{2} = \pm 1. 
\ea

Let us write down eqs. (2) in the variables $\zeta$ and $\eta$ \cite{MSh2023}:
\ba
\Big( \alpha  F_R (\ln \frac{\psi}{\alpha})_{\zeta}\Big)_{, \, \eta}  +   \Big( \alpha  F_R (\ln \frac{\psi}{\alpha})_{\eta}\Big)_{, \, \zeta} - \frac{ 2 F_R \,\psi^2}{\alpha} \, \omega_{, \,\zeta} \, \omega_{, \,\eta} & = & 0 \\
\Big(\frac{F_R \psi^{2}}{\alpha} \omega_{\zeta}\Big)_{, \, \eta}  + \Big(\frac{F_R \psi^{2}}{\alpha} \omega_{\eta}\Big)_{, \, \zeta}  & = & 0
\\
( \alpha F_R)_{, \,\, \zeta\eta} - \alpha f (F_R R - F) & = & 0 \\
\alpha  F_R \, f R + \alpha {F_R}_{, \,\, \zeta\eta} - 2 F_R \, {\alpha}_{, \,\, \zeta\eta} - \frac{1}{2} (\alpha_{, \,\, \eta} {F_R}_{, \,\, \zeta}  +  \alpha_{, \,\,\zeta} {F_R}_{, \,\,\eta}) & = & 0 
\ea
and two equations for metric coefficient $ f $ are:
\ba
(\ln f)_{, \,\,\zeta} = \dfrac{1}{\ln ( \alpha  F_R)_{, \,\, \zeta}} \Big(\ln ( \alpha  F_R)_{, \,\, \zeta \zeta} + \frac{\psi^2}{2\alpha^{2}} (\omega_{,\zeta})^{2} + \frac{1}{2} ((\ln \frac{\psi}{\alpha})_{,\zeta})^{2} + \frac{1}{2} ((\ln \alpha)_{,\zeta})^{2} + ((\ln F_R)_{,\zeta})^{2} \Big) 
\ea
and similar for $ (\ln f)_{, \,\,\eta} $. From this equation it is clear that we do not consider the cases with $ \alpha  F_R = const $. It is also useful to give the expression for $ (\ln f)_{, \,\zeta \,\eta} $:
\be
(\ln f)_{, \,\zeta \,\eta} = \dfrac{\alpha_{, \,\,\zeta} \alpha_{, \,\,\eta}}{\alpha^{2}} + \frac{1}{4} \textit{Tr} \, (g_{, \,\, \zeta} (g^{-1})_{, \,\, \eta}) - \frac{{F_{R}}_{, \,\, \zeta\eta}}{F_{R}} +  \dfrac{\alpha_{, \,\, \eta} {F_{R}}_{, \,\, \zeta}  +  \alpha_{, \,\,\zeta} {F_{R}}_{, \,\,\eta}}{2 \alpha  F_{R}}
\ee
although it is a consequence of the previous equations.

\section{Integration variables}

In this section we define the "travelling wave" variable $ y(\zeta, \eta) = \zeta + \lambda \eta $ that reduce system (5)-(9) to a system of ordinary differential equations. Since $\zeta$ and $\eta$ can be either one timelike and one spacelike coordinate, or both spacelike coordinates, it is necessary to specify type of $ \lambda $ so that $ y( \zeta, \eta ) $ be real. In \cite{MSh2023} two cases are examined and we present this result here: 
\be
y ( \zeta, \eta ) = \zeta + \lambda \eta =  \frac{1}{2} \big((\lambda + 1)z - (\lambda - 1) t \big)
\ee
and
\be
y( \zeta, \eta ) = \frac{1}{\sqrt{\lambda}} \,\,\,(\zeta + \lambda \eta) = \frac{1}{2\sqrt{\lambda}} \, \big((\lambda + 1)z - i \,\, (\lambda - 1) \rho \big). 
\ee

\section{Exact solutions}

Let us first consider eq. (6). In the variable $ y $ we immediately obtain 
\be
\omega_{y} = \frac{C_{0} \alpha }{\psi^{2} F_R} , \setcounter{equation}{13}
\ee
$ C_{0} = const $. From eq. (5) one can see that 
\be
\alpha F_R (\ln \frac{\psi}{\alpha})_{,y} - C_{0} \omega = C_{1},
\ee
$ C_{1} = const $ and from eqs. (5)-(6) we can obtain the expressions for $ \omega $ and $ F_R $:
\be
\omega = -\frac{C_{1}}{C_{0}} \pm \frac{\sqrt{C_{3}^{2} (\frac{\psi}{\alpha})^{2} - 1}}{\frac{\psi}{\alpha}}, \,\,\,
F_R = \pm C_0 \, \frac{\sqrt{C_{3}^{2} (\frac{\psi}{\alpha})^{2} - 1}}{\alpha  (\frac{\psi}{\alpha})_{,y}}
\ee
where $ C_{3}^{2} = (\frac{C_{1}}{C_{0}})^{2} + \frac{C_{2}}{C_{0}}$, $ C_{2} = const $. Let us now introduce the new function $ \chi(y) $ as:
\be
C_{3} \frac{\psi}{\alpha} = \cosh{\chi(y)},
\ee
$ \chi(y) \neq const $. Then 
\ba
\omega = -\frac{C_{1}}{C_{0}} \pm C_{3} \tanh \chi(y), \,\,\,\,
F_R = \frac{\varphi_{0}}{\alpha \chi_{,y}},
\ea
where $ \varphi_{0} =  \pm C_{0} C_{3} $ and you can see that $ \omega \neq 0 $.

Further we will express all functions of the model through $ \chi(y) $. Equation (7) can be rewritten as
\be
F = R \, F_R\, - \lambda \frac{( \alpha F_R)_{, \,\, y \, y}}{\alpha f}.
\ee
Differentiating this with respect to $ y $ and multiplying by $ f $ we obtain:
\be
f\,R\, (F_R)_y - \lambda \, \Big( \frac{( \alpha F_R)_{, \,\, y \, y}}{\alpha}\Big)_{, \,\, y} + \lambda \frac{( \alpha F_R)_{, \,\, y \, y}}{\alpha} \,\,(\ln f)_{, \,\, y} = 0.
\ee
Let us write out separately the terms in eq. (19). From eqs. (9)-(10):
\be
f\,R = \lambda \, \Big( \frac{( \alpha F_R)_{, \,\, y \, y}}{2 \alpha F_R} -  \frac{3 \alpha_{, \,\, y \, y}}{2 \alpha} -                                        \frac{{3  F_R}_{, \,\, y \, y}}{2 F_R} \Big)
\ee
and using eq. (17)
\be
f\,R\, (F_R)_y = \lambda \, \Big( \frac{\varphi_{0}}{\alpha \chi_{,y}}    \Big)_{y} \,\, \Big( \frac{\chi_{,\,yyy}}{\chi_{,\,y}} - \frac{2 (\chi_{,\,yy})^2}{(\chi_{,\,y})^2}  + \big( \frac{3 \alpha_{ ,\, y}}{\alpha} \big) _{y}  - \frac{3 \alpha_{ ,\, y}}{\alpha} \, \frac{\chi_{,\,yy}}{\chi_{,\,y}}\Big).
\ee
The second term in eq. (19) has the form:
\ba
\Big( \frac{( \alpha F_R)_{, \,\, y \, y}}{\alpha}\Big)_{, \,\, y} = - \frac{\lambda \varphi_{0}}{\alpha} \, \Big( \frac{\chi_{,\,yyyy}}{(\chi_{,\,y})^2} - \frac{6 \chi_{,\,yy}  \chi_{,\,yyy}}{(\chi_{,\,y})^3} + \frac{6  (\chi_{,\,yy})^3 }{(\chi_{,\,y})^4} - \frac{\alpha_{ ,\, y}}{\alpha} \big(  \frac{\chi_{,\,yyy}}{(\chi_{,\,y})^2} - \frac{2 (\chi_{,\,yy})^2}{(\chi_{,\,y})^3} \big) \Big);
\ea
third term is:
\be
\lambda \frac{( \alpha F_R)_{, \,\, y \, y}}{\alpha} \,\,(\ln f)_{, \,\, y} = - \frac{\lambda \varphi_{0}}{\alpha} \, \big(  \frac{\chi_{,\,yyy}}{(\chi_{,\,y})^2} - \frac{2 (\chi_{,\,yy})^2}{(\chi_{,\,y})^3} \big)  \,\,\, \Big( \frac{\chi_{,\,yyy}}{\chi_{,\,yy}} - \frac{2 \chi_{,\,yy}}{\chi_{,\,y}} - \frac{ (\chi_{,\,y})^3 }{2 \chi_{,\,yy}} - \big(\frac{\alpha_{ ,\, y}}{\alpha} \big)^2 \, \frac{ 3 \chi_{,\,y}}{2 \chi_{,\,yy}}  - \frac{2 \alpha_{ ,\, y}}{\alpha}  \Big).
\ee
Substituting eqs. (20)-(23) into eq. (19) we obtain the Abel differential equation of the second kind for the function $ z = (\ln \alpha)_{,\,y} $ ($ \alpha \neq const $):
\ba
z_{,\,y}\,\,(z + \frac{\chi_{,\,yy}}{\chi_{,\,y}}) & = & \frac{\chi_{,\,yyy}}{2 \chi_{,\,yy}} z^{2} + \big(\frac{\chi_{,\,yy}}{\chi_{,\,y}} \big)^{2} z + \nonumber \\ 
\frac{1}{3}\Big( \frac{\chi_{,\,yyyy}}{\chi_{,\,y}} & - & \frac{3 \chi_{,\,yyy} \chi_{,\,yy}}{(\chi_{,\,y})^{2}} +  \frac{4 (\chi_{,\,yy})^{3}}{(\chi_{,\,y})^{3}} - \frac{(\chi_{,\,yyy})^{2}}{\chi_{,\,y} \chi_{,\,yy}} +\frac{\chi_{,\,yyy}  (\chi_{,\,y})^{2}}{2\chi_{,\,yy}} - \chi_{,\,yy} \chi_{,\,y} \Big).
\ea
This equation can be solved exactly \cite{Kamke} and we obtain:
\be
\big(\ln (\alpha \chi_{,\,y})\big)_{,\,y} = \frac{1}{\sqrt{3}}  \Big[ \frac{2 \chi_{,\,yyy}}{\chi_{,\,y}} - \big(\frac{\chi_{,\,yy}}{\chi_{,\,y}} \big)^{2} - (\chi_{,\,y})^{2} \Big]^{\frac{1}{2}}.
\ee
From this it is clear that $ \chi $ must be chosen so that the expression under the square root is non-negative. In particular $\chi$ cannot have the form $\chi \sim y $ which leads to the exclusion of solutions to the model equations that give $ R = const $. 

Consider now eq. (9). In terms of the function ${\chi_{,\,y}} $ it has the form:
\be
(\ln f)_{,\,y} = \frac{\chi_{,\,yyy}}{\chi_{,\,yy}} - \frac{2 \chi_{,\,yy}}{\chi_{,\,y}} - \frac{(\chi_{,\,y})^3}{2 \chi_{,\,yy}} - \frac{3 \chi_{,\,y}}{2 \chi_{,\,yy}} \,\, \big( \frac{\alpha_{,\,y}}{\alpha} \big)^2 - \frac{2 \alpha_{,\,y}}{\alpha}.
\ee
Substituting $ \alpha $ from eq. (25) we obtain
\be
f = f_{0} \alpha,
\ee
where $ f_{0} = const $. Next substituting all expressions into eq. (8) we obtain an algebraic equation for determining $ R(y) $:
\be
R = \frac{\lambda}{f_{0} \alpha} \, \chi_{,\,y} \, \Big(  \big( \frac{2}{\chi_{,\,y}} \big)_{,\,yy}  + \big(  \frac{{\sqrt{3}} }{\chi_{,\,y}} \, \Big[ \frac{2 \chi_{,\,yyy}}{\chi_{,\,y}} - \big(\frac{\chi_{,\,yy}}{\chi_{,\,y}} \big)^{2} - (\chi_{,\,y})^{2} \Big]^{\frac{1}{2}}     \big)_{,\,y} \Big).
\ee
So it can be seen that all quantities included in eqs. (5)-(9) can be expressed through $ \chi(y) $ up to the point that an explicit form of $ F(y) $ can be obtained. Thus one can arbitrarily choose (taking into account the requirement of the reality of resulting solutions) the function $ \chi(y) $ and express everything in terms of $ y $. However, this does not bring us closer to the main goal - to find the analytical dependence of $ F(R) $ on $ R $. Moreover even the simplest choice of the function $ \chi(y) $ does not guarantee the existence of the inverse function $ y = y(R) $ from eq. (28). Therefore to further solve the model equations and find the exact form of the function $ F(R) $ we will make one important assumption: let function $ \chi_{,\,y} $ be function of $ R $: $ \chi_{,\,y} = \chi_{,\,y} (R) $. And for definiteness we will assume that $ R>0 $.

Because $ \alpha \chi_{,\,y} = \frac{\varphi_{0}}{F_R} $ (see eq. (17)) and $ F_R = F_R (R) $ is a function from $ R $ only we rewrite eq. (25) as:
\be
(\ln F_R)_{R} \, R_y = - \frac{1}{\sqrt{3}}  \Big[ \frac{2 \chi_{,\,yyy}}{\chi_{,\,y}} - \big(\frac{\chi_{,\,yy}}{\chi_{,\,y}} \big)^{2} - (\chi_{,\,y})^{2} \Big]^{\frac{1}{2}}.
\ee
Taking into account the assumption that $ \chi_{,\,y} = \chi_{,\,y} (R) $ eq. (29) becomes:
\be
2R_{,\,yy} + (R_{,\,y})^{2} \, \Big( 2 \frac{(\chi_{,\,y})_{,\,RR}}{({\chi_{,\,y}})_{,\,R}} -  \frac{(\chi_{,\,y})_{,\,R}}{{\chi_{,\,y}}} - \frac{ 3 \chi_{,\,y}}{({\chi_{,\,y}})_{,\,R}}\,\, \big( \frac{F_{RR}}{F_R} \big)^{2} \Big) - \frac{(\chi_{,\,y})^{3}}{({\chi_{,\,y}})_{,\,R}} = 0. 
\ee
Solving this equation we obtain:
\be
(R_{,\,y})^{2} =  \frac{\chi_{,\,y}}{(({\chi_{,\,y}})_{,\,R})^2} \,\,e^{3 \int \frac{\chi_{,\,y}}{({\chi_{,\,y}})_{,\,R}} \, \big( \frac{F_{RR}}{F_R} \big)^{2} dR} \, \Big( \int ({\chi_{,\,y}})^{2} \,\, ({\chi_{,\,y}})_{,\,R}\,e^{-3 \int \frac{\chi_{,\,y}}{({\chi_{,\,y}})_{,\,R}} \, \big( \frac{F_{RR}}{F_R} \big)^{2} dR} dR +  p_{0}  \Big),
\ee
$ p_{0} = const $, and the dependence $y(R) $ is
\be
y+y_0 = \pm \int \frac{dR}{\sqrt{ \frac{\chi_{,\,y}}{(({\chi_{,\,y}})_{,\,R})^2} \,\,e^{3 \int \frac{\chi_{,\,y}}{({\chi_{,\,y}})_{,\,R}} \, \big( \frac{F_{RR}}{F_R} \big)^{2} dR} \, \Big( \int ({\chi_{,\,y}})^{2} \,\, ({\chi_{,\,y}})_{,\,R}\,e^{-3 \int \frac{\chi_{,\,y}}{({\chi_{,\,y}})_{,\,R}} \, \big( \frac{F_{RR}}{F_R} \big)^{2} dR} dR +  p_{0} \Big)}}.  
\ee
This equation shows that our scheme assumes that function $ y(R) $ can be obtained at least in quadratures. However if the integral in eq. (32) is not taken or it is not possible to obtain $ R $ as $ R(y) $ then it is impossible to obtain an explicit dependence of the metric on $ y $. 

And finally let us consider eq. (8) again. The algebraic eq.(28) turns into a differential equation on $ R(y) $:  
\be
R = \frac{\lambda}{\varphi_{0} f_{0}}  F_R \, \Big( -2 \chi_{,\, yyy} + \frac{4 (\chi_{, \,yy})^{2}}{\chi_{,\,y}} - 3 \chi_{,\,y} \big( (\ln F_R)_{RR} (R_{,\,y})^{2} + (\ln F_R)_{R} R_{,\,yy} \big) + 3 (\ln F_R)_{R} R_{,\,y} \chi_{,\, yy} \Big).
\ee
Given that $ \chi_{,\,y} = \chi_{,\,y} (R) $ and $ F_R = F_R(R) $, introducing the notation $N(R) =  - \frac{2 (\chi_{,\, y})_R}{(\chi_{,\, y})^2} - \frac{3 (\ln F_R)_{R}}{\chi_{,\, y}} $ (we require that $ N \neq 0 $) eq. (33) will take the form:
\be
R_{,\,yy} + \big(\ln N \big)_{R} (R_{,\,y})^2 - \frac{\varphi_{0} f_{0}}{\lambda}\, \frac{R}{F_R \, N (\chi_{,\, y} )^{2}} = 0.
\ee
By integrating we obtain 
\be
(R_{,y})^{2} = \frac{1}{N^{2}} \Big( \frac{2 \varphi_{0} f_{0}}{\lambda} \int \frac{R N dR}{F_R (\chi_{,\, y})^2}  + N_0 \Big), \,\,\,\, N_0 = const.
\ee

Equating $(R_{,y})^{2}$ from eqs. (31) and (35)
\ba
\frac{\chi_{,\,y}}{(({\chi_{,\,y}})_{,\,R})^2} \,\,e^{3 \int \frac{\chi_{,\,y}}{({\chi_{,\,y}})_{,\,R}} \, \big( \frac{F_{RR}}{F_R} \big)^{2} dR} \, \Big( \int ({\chi_{,\,y}})^{2} \,\, ({\chi_{,\,y}})_{,\,R}\,e^{-3 \int \frac{\chi_{,\,y}}{({\chi_{,\,y}})_{,\,R}} \, \big( \frac{F_{RR}}{F_R} \big)^{2} dR} dR  +  p_{0} \Big) = \\
 \frac{1}{N^{2}} \Big( \frac{2 \varphi_{0} f_{0}}{\lambda} \int \frac{R N dR}{F_R (\chi_{,\, y})^2}  + N_0 \Big) \nonumber
\ea
we obtain the equation for determining the function $ F_R $ as $ F_R = F_R (\chi_{,\,y}(R)) $ with known $ \chi_{,\,y}(R) $ (or vice versa). But this is not the only possibility to solve this equation; also, for example, one can put $ \chi_{,\,y}$ as a function of $ F_R $ and express them in terms of $ R $, which is independently included in the resulting equation.

However, despite the apparent simplicity of this scheme its practical implementation encounters difficulties. As can be seen from eq. (36) this expression contains three integrals, one of which cannot be removed simply by differentiation. Nevertheless it seems to us that in general the first step in solving eq. (36) may be to multiply it by $ N^{2} $ and then differentiate it with respect to $ R $. In this case the constant $N_0$ (or, possibly, $p_{0}$) obviously disappears, so to make it more specific subsequent substitution of the resulting solutions into the initial equations for $ R_y $ is necessary. And then having solved this equation and obtained an exact expression for the function $ F_R $ we restore $F(R) $ as $ F(R) =  \int F_R dR + F_0$. Thus to obtain the exact form of $ F(R) $ there is no need to solve the initial eq. (7) that is much more complicated. However, as for the constant $N_0$ here one will have to substitute the resulting function $F(R) $ in eq. (7) to determine $ F_0 $.

As was seen from eq. (17) we do not get a diagonal metric since the off-diagonal component of $ g_{ab} $ cannot be equal to zero. Let us now consider the diagonal metric: $ g_{ab} = \textit{diag}(\psi, \alpha^{2}\, \psi^{-1} )$. Eq. (6) is satisfied automatically and eq. (5) gives:
\be
F_R = \frac{C_1}{\alpha  (\ln \frac{\psi}{\alpha})_{,y}}.
\ee
If we now introduce the function $ \chi(y) $ as
\be
\chi(y) = \ln \frac{\psi}{\alpha}
\ee
which is different from eq. (16) we obtain  
\be
F_R = \frac{C_1}{\alpha \chi_{,y}}
\ee
which coincides with the second equation of eq. (17). Next from eq. (9) on can see that the expression for $ (\ln f)_{, y} $ completely coincides with eq. (26) and we can conclude that the further scheme for finding $ F(R) $ is identical to the scheme described above. Thus the same $ \chi(y) $ and the same $ F(R) $ correspond to two metrics, diagonal and non-diagonal, and one does not transform into the other. The metric interval (3) in the variables $ \frac{1}{2} (z \pm t) $ for $ \lambda = -1 $ in eq. (11) will be written as
\be
ds^{2} =  \alpha \, dt^2 - \alpha \, ( dz^{2} + \frac{\psi}{\alpha} dx^2 + \frac{\alpha}{\psi} dy^2)
\ee
and it can be seen that the case $ \psi \sim \alpha $, for example FLRW metric, cannot be considered since at $ \psi \sim \alpha $ the proposed integration scheme does not work.

In the next section we consider some examples of the application of the above scheme.

\section{Some examples}

\subsection{}

First let us consider the case that perhaps first comes to mind when we see the integral in the exponent. Let $ \chi_{,\,y} = u _0 (F_R)^{k} $, $ k \neq 0, \pm 1, - \frac{3}{2}, -3 $. Substituting this in eq. (36):
\be
\frac{(2k+3)^2}{{u_0}^3 k^2} \, \Big( \frac{{u_0}^3 k^2}{3k^2-3} +  p_{0} \, (F_R)^{-3k+\frac{3}{k}}   \Big) = - \frac{2 \varphi_{0} f_{0} (2k+3)}{\lambda {u_0}^3} \int R F_{RR} \, (F_R)^{-3k-2} dR  + N_0. 
\ee
Differentiating this with respect to $ R $ we obtain:
\be
F_R = \phi_0 \, R^{\frac{k}{k+3}}, \,\,\, \phi_0  =  \Big(  \frac{2 \varphi_{0} f_{0} k^3 }{\lambda p_0 (2k+3) (3k^2-3)   }  \Big)^{\frac{k}{k+3}} .
\ee
Integrating over $R$ gives:
\be
F(R) = \frac{k+3}{2k+3}\,\phi_0 \,\, R^{\frac{2k+3}{k+3}},
\ee
where $ n \neq 1 $ and $ n \neq 2 $. The case $ F(R) \sim R^n $ was considered in \cite{MSh2023} and for some values of $ n $ exact solutions were obtained for all model variables. Substituting the resulting solution into eq. (36) we find that $ N_0 = \frac{(2k+3)^2}{3k^2-3} $. Now let us find the function $ y(R) $ from eq. (31): 
\be
y + y_0 = C_4 \, R^{\frac{k^2-3}{2(k+3)}}\,_{2}F_{1}\big(\frac{1}{2},\,\, \frac{k^2-3}{6 k^2-6};\,\, \frac{7k^2-9 }{6k^2-6}; \,\, C_5 R^\frac{3k^2-3}{k+3} \big),
\ee
$ C_4 = \pm \frac{2k^2}{k^2-3} \sqrt{\frac{u_0}{p_0}} {\phi_0}^{\frac{3-k^2}{k}}  $, where the $ + $ sign is taken for $ k \in (-3, 0) $ and the $ - $ sign is taken for $ k \in (-\infty, -3) \cup (0, \infty) $; $ C_5 = - \frac{{u_0}^3 k^2}{p_0 (3k^2-3)}  {\phi_0}^{\frac{3k^2-3}{k}}$. It can be seen that solution (44) is not applicable when $ k = \pm \sqrt{3} $. This case was analyzed in \cite{MSh2023} where an exact expression for $ R(y) $ was obtained. In the case of arbitrary $ k $ it is impossible to express $ R $ as $ R(y) $; the values of $ k $ when this can be done are discussed in the work mentioned above. 

Thus we obtain that for arbitrary $ k $ all metric functions are expressed in terms of $ R $ and they have the form:
\ba
\alpha(R) & = & \frac{\varphi_0 }{u_0 {\phi_0}^{k+1}}\,\,R^{-\frac{k(k+1)}{k+3}}  \\
\chi(R)& = & \pm \frac{2k}{\sqrt{3k^2-3}} \,\, \sinh^{-1} \Big( \sqrt{\frac{{u_0}^3 k^2 {\phi_0}^{\frac{3k^2-3}{k}}}{p_0}} \,\,\frac{R^{\frac{3k^2-3}{2(k+3)}}}{\sqrt{3k^2-3}}   \Big) + \chi_0, \,\,\,\, 3k^2-3 > 0 \nonumber \\
\chi(R)& = & \mp \frac{2k}{\sqrt{3-3k^2}} \,\, \arcsin \Big( \sqrt{\frac{{u_0}^3 k^2 {\phi_0}^{\frac{3k^2-3}{k}}}{p_0}} \,\,\frac{R^{\frac{3k^2-3}{2(k+3)}}}{\sqrt{3-3k^2}}   \Big) + \chi_0,, \,\,\,\, 3k^2-3 < 0,\nonumber
\ea
where the signs $+$ and $-$ are taken as in $ C_4 $. The remaining functions $ \psi $, $ \omega $ and $ f $ are expressed in terms of $ \chi(R) $ from eqs. (16), (17) and (27) respectively. 

And at the end of this section we consider the values that $ k $ cannot take. At $ k = 0 $ the function $ \chi_{,\,y} = const $ and we do not consider this, see eq. (25). When $ k = - \frac{3}{2} $, $ N = 0 $ and eq. (34) gives $ R = 0 $; for $ k = - 3 $ eq. (41) reduces to $ R = const $. It is of interest to study the case $ k = \pm 1 $ in more detail. If we consider eq. (7) in the form
\ba
F(R) & = & R F_R - \frac{\lambda}{\varphi _0  f_0} \, {F_R}^2 \, \Big( - \frac{1}{2} ({\chi_{,\,y}})^{3}  + \\
& + &  \frac{3}{2} \big( 1 - \big( \frac{{\chi_{,\,y}}}{({\chi_{,\,y}})_{, \,R}} \,\,\frac{F_{RR}}{F_R}  \big)^{2}  \big) \,\,e^{3 \int \frac{\chi_{,\,y}}{({\chi_{,\,y}})_{,\,R}} \, \big( \frac{F_{RR}}{F_R} \big)^{2} dR} \, \Big( \int ({\chi_{,\,y}})^{2} \,\, ({\chi_{,\,y}})_{,\,R}\,e^{-3 \int \frac{\chi_{,\,y}}{({\chi_{,\,y}})_{,\,R}} \, \big( \frac{F_{RR}}{F_R} \big)^{2} dR} dR  +  p_{0} \Big)        \Big) \nonumber
\ea
we can see that  $ k = \pm 1 $ it simplifies significantly and solutions can be easily found using the ansatz $ F(R)\sim R^n $. We do not consider the formal solution obtained for both $k$ of the form $F(R) = A\, R+B$ since $  n \neq 1 $ and $ R \neq const $. Other solutions have the form: for $ k = - 1 $ the function $ F(R)\sim \sqrt{R} $ but $ \alpha = const $ and we are not study such a case here (see eq. (24)). For $ k=1 $ we obtain:
\ba
F(R) & = & \frac{4}{5} \, \big( - \frac{2 \varphi_0 f_0 }{5 \lambda {u_0}^3}  \big)^{\frac{1}{4}} \,\, R^{\frac{5}{4}} \nonumber \\
y + y_0 & = & \big( - \frac{5 \lambda {u_0}}{2 \varphi_0 f_0}\big)^{\frac{1}{4}} \, e^{\frac{N_0}{25}} \, \sqrt{\pi} \, erf \big( \frac{1}{2} \sqrt{\frac{4 N_0}{25} + \ln R}  \big),
\ea
$ N_0 = \frac{25 p_0}{{u_0}^3} + \frac{25}{4} \ln \big( - \frac{2 \varphi_0 f_0 }{5 \lambda {u_0}^3}\big) $ and it is obvious that again it cannot be found $ R = R(y) $. In fact these solutions (with $ k = \pm 1 $) do not represent anything radically new and were not included in \cite{MSh2023} precisely because of the impossibility of expressing model variables as functions of $ y $.

\subsection{}

Considering $ F(R) \sim R^n $ we excluded the case $ n = 2 $. Let us now obtain the exact solution for all metric functions in the case $  F(R) = F_0 \, R^2$. Let
\be
F_R = \phi_0 \, \Big( ({\chi_{,\,y}})^{\frac{3}{2}} + \sqrt{({\chi_{,\,y}})^{3} + a}  \Big)^{\frac{2 \sqrt{u_0}}{\sqrt{3}}},
\ee
where $ \phi_0 $, $ a $ and $ u_0 $ are arbitrary non-zero constants, $ p_0 = 0 $. Solving eq. (36) we obtain 
\be
{\chi_{,\,y}} = \Big( \frac{1}{2} \big( \frac{R}{\delta} \big)^{\frac{\sqrt{3}}{2 \sqrt{u_0}}} -\frac{a}{2} \big( \frac{R}{\delta} \big)^{- \frac{\sqrt{3}}{2 \sqrt{u_0}}}  \Big)^{\frac{2}{3}}, 
\ee
$ \delta = \frac{a \lambda \phi_0}{\varphi_0 f_0 (1-3 u_0)} $, $ u_0 \neq \frac{1}{3} $ and we immediately obtain $ F_R =  \frac{\phi_0 R}{\delta} $ which gives
\be
F(R) = \frac{\phi_0}{2\delta} \, R^2.
\ee
The constant $ N_0 $ is equal to $ N_0 = 27 u_0 + 4 $. Expressing $ y(R) $ from eq. (35):
\be
y + y_0 =  \pm 2^{\frac{2}{3}}  \sqrt{3(1 - 3u_0)}  \big( \frac{R}{\delta} \big)^{- \frac{1}{\sqrt{3 u_0}}}  \,_{2}F_{1}\Big(\frac{1}{3},\,\, \frac{2}{3};\,\, \frac{4 }{3}; \,\, a \, \big( \frac{R}{\delta} \big)^{- \sqrt{\frac{3}{u_0}}}\Big),
\ee
which again makes it impossible to express $ R $ as a function of $ y $. The same can be said about other functions:
\ba
\chi(R) & = & \sqrt{\frac{1-3 u_0}{u_0}} \, \ln R + \chi_0, \,\,\,\,\, \chi_0 = const  \nonumber \\
\alpha(R)  & = & \frac{\varphi_0  \delta}{\phi_0} \,\, \frac{1}{R }\,\,  \Big( \frac{1}{2} \big( \frac{R}{\delta} \big)^{\frac{\sqrt{3}}{2 \sqrt{u_0}}} -\frac{a}{2} \big( \frac{R}{\delta} \big)^{ - \frac{\sqrt{3}}{2 \sqrt{u_0}}}  \Big)^{-\frac{2}{3}} \\
\psi(R) & = & \frac{\alpha}{C_3} \,\, \cosh \Big( \sqrt{\frac{1-3 u_0}{u_0}} \, \ln R + \chi_0 \Big) \nonumber \\
\omega(R) & = & -\frac{C_{1}}{C_{0}} \pm C_{3} \tanh \Big( \sqrt{\frac{1-3 u_0}{u_0}} \, \ln R + \chi_0 \Big) \nonumber
\ea
and $ f = f_0 \alpha $.


Now let us consider this model $ F(R) = F_0 R^2 $ in more detail. It is interesting that this model allows us to integrate it without resorting to the assumption that $ \chi_{,\,y} = \chi_{,\,y}(R) $. And it also gives a simple example of where our integration scheme does not work.

The combination of eqs. (7) and (27) gives 
\be
(\alpha \, R)_{\, y \, y } = \frac{f_0}{2 \lambda} (\alpha \, R)^2
\ee
and together with eq. (8) which can be rewritten as
\be
\frac{\alpha_{\, y \, y}}{\alpha} - \frac{R_{\, y \, y}}{R}  =  \frac{f_0}{2 \lambda} \alpha \, R
\ee
this leads to the relationship
\be
\alpha \, R_{\, y } = D_0,
\ee
where $ D_0 = const $. Substituting $ \alpha = \frac{D_0}{ R_{\, y }} $ in eq. (53) we obtain
\be
\Big( \big( \frac{R}{R_{\, y}} \big)_{\, y} \Big)^2 = \frac{f_0 D_0}{2 \lambda} \, \Big( \frac{R}{R_{\, y}} \Big)^{3} + D_1,
\ee
$ D_1 = const $. Moving on to $ P = (R_{\, ,y})^3 $ this equation becomes
\be
R^{2} \,\, (P_{\, ,R})^2 - 6 R P P_{\, ,R} - \frac{3 f_0 D_0}{\lambda} R^3 P + (1 - D_1) P^2 = 0. 
\ee
Solving this equation and returning to $ R $ we get:
\be
y + y_0 =  \, \Big( \frac{12 \lambda b  D_1}{ f_0 D_0} \Big)^{\frac{1}{3}} \,   \frac{1}{\sqrt{D_1}} \, R^{- \sqrt{D_1}} \, _{2}F_{1}\Big(\frac{1}{3},\,\, \frac{2}{3};\,\, \frac{4 }{3}; \,\, - \frac{b}{R^{ 3 \sqrt{D_1}}} \Big),
\ee
where $ b $ is a positive constant, $ D_1 > 0$. The comparison of expressions (51) and (58) shows that (58) is more general; setting $ D_1 = 1 $ we obtain
\be
y + y_0 =   \Big( \frac{12 \lambda b}{ f_0 D_0} \Big)^{\frac{1}{3}} \, \frac{1}{R} \, _{2}F_{1}\Big(\frac{1}{3},\,\, \frac{2}{3};\,\, \frac{4 }{3}; \,\, - \frac{b}{R^3} \Big)
\ee
but $ \sqrt{\frac{3}{u_0}} \neq 3 $ ($ u_0 \neq \frac{1}{3} $). The function $  \alpha $ has the form:
\be
\alpha (R) =  \Big(- \frac{12 \lambda D_1 {D_0}^2}{ f_0 } \Big)^{\frac{1}{3}} \, \frac{1}{R} \, \Big( b R^{\frac{3 \sqrt{D_1}}{2}} + b^
{-1} R^{- \frac{3 \sqrt{D_1}}{2}}  \Big)^{- \frac{2}{3}}
\ee
and also differs from eq. (52) by the solution with $ D_1 = 1 $. The remaining functions have the form:
\ba
\chi(R) & = & \frac{{C_0} C_3 }{2 {F_0} D_0} \, \ln R + \chi_1, \,\,\,\,\, \chi_1 = const  \nonumber \\
\psi(R) & = & \frac{\alpha}{C_3} \,\, \cosh \Big( \frac{{C_0} C_3 }{2 {F_0} D_0} \, \ln R + \chi_1 \Big) \nonumber \\
\omega(R) & = & -\frac{C_{1}}{C_{0}} \pm C_{3} \tanh \Big( \frac{{C_0} C_3 }{2 {F_0} D_0} \, \ln R + \chi_1 \Big) 
\ea

Let us now consider the case $ D_1 = 0 $ and show a situation where our integration scheme cannot be applied. Using eqs. (55) and (56) we find
\ba
R(y) & = & R_0 \, e^{\frac{f_0 D_0}{36 \lambda} \, (y + y_0)^3} \nonumber \\
\alpha (y) & = & \frac{12 \lambda}{R_0 f_0} \, e^{-\frac{f_0 D_0}{36 \lambda} \, (y + y_0)^3} \, (y - y_0)^{-2},
\ea
$ y_0 = const$. Then from the second equation (17) (and eq. (39) we obtain that
\be
\chi_{\, ,y} (y) = \chi_0 (y - y_0)^{2}
\ee
and one can see that this function is not suitable for our scheme since then the radical expression in eq. (25) becomes negative. In other words if one start with the choice of $ \chi $ then the fact that this function does not correspond to the domain of definition is established immediately. Confirm this as follows. This also affects the expression for $ f(y) $; so for the diagonal metric from eq. (26):
\be
f(y) = f_0 \, \alpha \, e^{- \frac{1}{24} ({\chi_0}^2 + \frac{{f_0}^2 {D_0}^2}{48 \lambda^2}) (y+y_0)^6   }
\ee
and one can see that eq. (27) is not satisfied; it would be satisfied if $ \chi_0 = i |\chi_0| $. The resulting contradiction means that eqs. (53)-(55) also turn out to be incorrect in this case and expressions (62) we found are not solutions. Repeat that although it is formally possible to obtain the expression $ \chi_{\, ,y} =  \chi_{\, ,y} (R) $ we cannot use it in our scheme due to non-compliance with the domain of definition of eq. (25).

Let us now turn to the work \cite{IV2021} and obtain exact solutions to the
corresponding GTR model with a constant potential using the solutions (60)-(61): 
\be
S = - \frac{1}{16 \pi k}  \int d^{4}x \sqrt{-\tilde{\textit{g}} }\,\Big( \tilde{R} - \frac{1}{2} \tilde{g^{\mu \nu }}  \phi_{,\mu} \phi_{, \nu} - \Lambda \Big),  
\ee
where $ \phi $ is a scalar field, $ \Lambda $ is a constant potential. Using the conformal transformation of the metric:
\be
\tilde{g^{\mu \nu }} = K_0 e^{\kappa \phi} \, {g^{\mu \nu }}
\ee
we obtain that 
\be
\tilde{R} = K_0 \, e^{\kappa \phi } \,{R} + 3 K_0 \, \kappa \, e^{\kappa \phi }
{g^{\mu \nu }} {\bigtriangledown}_{ \mu} {\phi }_{, \nu} - \frac{3}{2} \, K_0 \, \kappa^{2} \, e^{\kappa \phi } \, {g^{\mu \nu }} \phi_{,\mu} \phi_{, \nu}.
\ee
Substituting this into the Lagrangian (removing the complete divergence):
\be
S = - \frac{1}{16 \pi k}  \int d^{4}x \sqrt{-{\textit{g}}} \,\Big( 
\frac{e^{- \kappa \phi }}{K_0} {R} + \frac{3 \kappa^{2} - 1}{2 K_0} {g^{\mu \nu }} \phi_{,\mu} \phi_{, \nu} - \frac{e^{- 2 \kappa \phi }}{{K_0}^2} \,\, \Lambda \Big).  
\ee
If we put $ \kappa = \pm \frac{1}{\sqrt{3}} $ the second term will disappear. Then varying action over $ \phi $
\be
e^{\mp \frac{\phi}{\sqrt{3}}} = \frac{K_0}{2 \Lambda} R
\ee
and substituting this into this action we get \cite{IV2021}:
\be
S = - \frac{1}{16 \pi k}  \int d^{4}x \sqrt{-{\textit{g}}} \, \frac{1}{4 \Lambda} \,\,{R}^{2}.  
\ee
Thus using the obtained solutions (60)-(61) for $ F(R) = F_0 R^2 $ we can find solutions of the GTR model with a scalar field and a constant potential expressed through the Ricci scalar. For the variable $ y $ let us take for definiteness eq. (11). Substituting eq. (69) into eq. (67) we obtain:
\be
\tilde{R} = 2 \Lambda + \frac{12 \Lambda \lambda }{f R }\,\, \Big(  \big( \frac{R_y}{R}  \big)^2 + 2 \, \big( \frac{R_y}{R}  \big)_{\, ,y} + 2 \,\frac{R_y}{R} \,\,\frac{\alpha_{\,,y}}{\alpha}  \Big)
\ee
which using eqs. (59) and (60) gives the following expression for the scalar curvature:
\be
\tilde{R} = 2 \, \Lambda + \frac{\Lambda}{D_1} \,\, \Big( b R^{\frac{3 \sqrt{D_1}}{2}} + b^{-1} R^{- \frac{3 \sqrt{D_1}}{2}}  \Big)^{2}.
\ee
From eq. (66) we obtain that $ \tilde{g_{\mu \nu }} = \frac{R}{2 \Lambda} \, {g_{\mu \nu }} $ where $ g_{\mu \nu } $ is defined by eqs. (3), (4) (with $ \sigma_{\alpha}^{2} = + 1 $), (60) and (61).

\subsection{}

Now by analogy with eq. (48) consider the function $ F_R $ of the form:
\be
F_R = \phi_0 \, \Big( ({\chi_{,\,y}})^{\frac{3}{4}} + \sqrt{({\chi_{,\,y}})^{\frac{3}{2}} + a}  \Big)^{\pm 1}
\ee
where $ u_0 = \frac{3}{8} $, $ a \neq 0 $, $ p_0 = 0 $. Let us denote $ ({\chi_{,\,y}})^{\frac{3}{4}} \pm \sqrt{({\chi_{,\,y}})^{\frac{3}{2}} + a} \equiv g_{\pm} $ and then we will write all the results in terms of $ g_{\pm} $. Substituting eq. (73) in eq. (36) and differentiating it with respect to $ R $ we obtain cubic equations for $ g_{\pm} $ whose solutions are:
\be
g_{\pm} = \Big( \kappa_{\pm} \, R + \sqrt{\big( \kappa_{\pm} \big)^{2} \,\, R^2 + \big( \frac{23 a}{27} \big)^{3}}\Big)^{\frac{1}{3}} + \Big( \kappa_{\pm} \, R - \sqrt{\big( \kappa_{\pm} \big)^{2} \,\, R^2 + \big( \frac{23 a}{27} \big)^{3}}\Big)^{\frac{1}{3}},
\ee
$ \kappa_{+} = \frac{7\varphi_0 f_0}{72 \lambda a \phi_0}$, $  \kappa_{-} = - a  \kappa_{+}$ and $ {\chi_{,\,y}} = \big( \frac{g^2-a}{2g} \big)^{\frac{4}{3}} $, $ N_0 = \frac{289}{21} $. Substituting this into eq. (73) we obtain (for $ g_{+} $):
\ba
F(R) & = & \frac{3 \phi_0}{4} \, R \, g - \frac{23 \lambda a^2 {\phi_0}^2}{7\varphi_0 f_0} \,\,\Big[ \Big( \frac{7\varphi_0 f_0}{72 \lambda a \phi_0} \, R + \sqrt{\big( \frac{7\varphi_0 f_0}{72 \lambda a \phi_0} \big)^{2} \,\, R^2 + \big( \frac{23 a}{27} \big)^{3}}\Big)^{\frac{2}{3}} \nonumber \\
& + & \Big( \frac{7\varphi_0 f_0}{72 \lambda a \phi_0} \, R - \sqrt{\big( \frac{7\varphi_0 f_0}{72 \lambda a \phi_0} \big)^{2} \,\, R^2 + \big( \frac{23 a}{27} \big)^{3}}\Big)^{\frac{2}{3}} \Big] + \frac{815 \lambda a^3 {\phi_0}^2}{189 \varphi_0 f_0}. 
\ea
But here the situation with the ability to express $ y(R) $ functionally has become worse; $ y $ is expressed only in quadratures:
\be
y + y_0 = \pm \sqrt{\frac{28}{3}} \, \int \Big( \frac{2g}{g^2-a}\Big)^{\frac{1}{3}} \, \Big( g^4 + 30ag^2 + a^2  \Big)^{- \frac{1}{2}} dg.
\ee
The expression for $ \chi(R) $ is obtained exactly and has the form:
\be
\chi(R) = \pm \frac{\sqrt{7}}{4\sqrt{3}} \ln \Big( \frac{g^4 + 16 a g^2 + a^2 + (g^2 +a)\sqrt{g^4 + 30ag^2 + a^2}}{g^4 + 16 a g^2 + a^2 - (g^2 +a)\sqrt{g^4 + 30ag^2 + a^2}}  \Big) + \chi_0
\ee
and metric functions will be expressed through $ \chi(R) $ using the formulas (15)-(17) and (27).

\subsection{}

The following example shows that it is not always possible to fully integrate the problem, and it is possible to obtain $ F(R) $ only as an integral. Consider the function $ {\chi_{,\,y}} $ in the form:
\be
{\chi_{,\,y}} = \chi_0 \, \Big( (F_R)^{2} + a (F_R)^{-\frac{3}{2}} \Big)
\ee
where $ \chi_0 = const $, $ a \neq 0 $, $ p_0 = 0 $. Substituting eq. (78) in eq. (36) and differentiating it with respect to $ R $ we obtain the cubic equation on $ (F_R)^{-\frac{7}{2}} \equiv g $:
\ba
\big( 21 a^3 - \frac{9 \varphi_0 f_0 a^2}{56 \lambda {\chi_0}^{3}}\, R \big) \, g^3 +  \big(- \frac{7 a^2(9 a^2 + 35)}{10} + \frac{12 \varphi_0 f_0 a}{7 \lambda {\chi_0}^{3}}\, R \big) \, g^2 + \nonumber \\ 
\big( \frac{14 a(10 + 5a + 2 a^2)}{5} - \frac{8 \varphi_0 f_0}{7 \lambda {\chi_0}^{3}}\, R \big) \, g - \frac{7a (14a+15)}{10} =0
\ea
whose solutions are:
\ba
g = \frac{1}{3} \big( \frac{7 a^2(9 a^2 + 35)}{10} - \frac{12 \varphi_0 f_0 a}{7 \lambda {\chi_0}^{3}}\, R \big) \, \big( 21 a^3 - \frac{9 \varphi_0 f_0 a^2}{56 \lambda {\chi_0}^{3}}\, R \big)^{-1} \, + \zeta_{+} + \zeta_{-}  \nonumber \\
\zeta_{\pm} = \big( 21 a^3 - \frac{9 \varphi_0 f_0 a^2}{56 \lambda {\chi_0}^{3}}\, R \big)^{-1} \, \Big( \gamma_{0}+ \gamma_{1} R + +\gamma_{2} R^2 + \gamma_{3} R^3  \pm \sqrt{{\Sigma^{6}_{i=0}} \beta_{i}R^i} \Big)^{\frac{1}{3}}
\ea
where $ \beta_i $ and $ \gamma_i $ are constants (depending on $ a$ and other constants of the model). As can be seen from this expression $ F(R) $ can only be obtained in quadratures:
\be
F(R)  =  \int g^{-\frac{2}{7}}dR,
\ee
where the integral cannot be taken exactly. The same can be said about 
$$ y(R) \sim \int g^{-\frac{3}{7}} \sqrt{\frac{1+\frac{3a}{4}g}{(1+ag)(\frac{2}{3}+2ag-\frac{2a^2}{5}g^2})} dg $$. This example shows that even finding an exact solution of eq. (36) does not guarantee the obtaining of solution for $ F(R) $ in integrated form.

\section{Conclusion}

In this article we present a scheme for obtaining exact analytical solutions for the metric functions in $F(R)$ gravity and for the function $F$ itself in terms of the travelling wave variable. Since it is important for us to obtain an explicit dependence of the function $F$ on the scalar curvature, we imposed the condition that all variables of the theory depend on $R$. The resulting components of the metric tensor also turn out to depend on the scalar curvature which complicates their physical analysis. However, in this article we digress from the analysis of the properties of the objects described by our solutions and focus specifically on the possibility of obtaining different forms of the function $F(R)$. 

For case $ F(R) \sim R^2 $ we also find the exact solution by direct integration and it turns out to be almost identical to that obtained after applying the above scheme. Also referring to the results of the work \cite{IV2021} and using the obtained solutions for $ F(R) \sim R^2 $ we expressed the GTR model with a scalar field and a constant potential in terms of the Ricci scalar of $ F(R) \sim R^2 $ model.

This article assumes further development in several directions. One can obtain different exact types of the function $F$. Also it is possible to analyse the solutions even if $F$ is obtained in quadratures, for example, by expanding the integrand into series rather than by expanding in the original equations, that simplifies obtaining solutions to the equations in the form of series. 

We find it interesting to apply the described scheme to find exact solutions of GTR with a cosmological constant; this work is in progress.

\end{document}